\documentclass[fleqn,10pt]{wlscirep}
\usepackage[utf8]{inputenc}
\usepackage[T1]{fontenc}

\usepackage{graphicx,subfig,float}
\usepackage{dcolumn}
\usepackage{bm}
\usepackage[normalem]{ulem}
\usepackage{xcolor}

\graphicspath{{./figures/}} 

\newlength{\subcolumnwidth}

\newcommand{\nextsubcolumn}[1][]{%
  \cr\noalign{\hfill}
  \if\relax\detokenize{#1}\relax\else\hsize=#1\setlength{\subcolumnwidth}{\hsize}\fi
}

\usepackage{lineno}

\title{Wide-field magnetometry using nitrogen-vacancy color centers with randomly oriented micro-diamonds}

\author[1,*]{Saravanan Sengottuvel}
\author[1]{Mariusz Mr\'{o}zek}
\author[2]{Miros{\l}aw Sawczak}
\author[3]{Maciej ~J. G{\l}owacki}
\author[3]{Mateusz Ficek}
\author[1]{Wojciech Gawlik}
\author[1,*]{Adam ~M. Wojciechowski}
\affil[1]{Institute of Physics, Jagiellonian University in Krakow, 11 {\L}ojasiewicza St., 30-348 Krak\'{o}w, Poland}
\affil[2]{Szewalski Institute of Fluid-Flow Machinery, Polish Academy of Sciences, 14 Fiszera St., 80-231 Gda\'nsk, Poland}
\affil[3]{Gda\'nsk University of Technology, 11/12 G. Narutowicza St., 80-233 Gda\'nsk, Poland}

\affil[*]{saravanan.sengottuvel@doctoral.uj.edu.pl, a.wojciechowski@uj.edu.pl}

\begin{abstract}
Magnetometry with nitrogen-vacancy (NV) color centers in diamond has gained significant interest among researchers in recent years. Absolute knowledge of the three-dimensional orientation of the magnetic field is necessary for many applications. Conventional magnetometry measurements are usually performed with NV ensembles in a bulk diamond with a thin NV layer or a scanning probe in the form of a diamond tip, which requires a smooth sample surface and proximity of the probing device, often limiting the sensing capabilities. Our approach is to use micro- and nano-diamonds for wide-field detection and mapping of the magnetic field. In this study, we show that NV color centers in randomly oriented submicrometer-sized diamond powder deposited in a thin layer on a planar surface can be used to detect the magnetic field. Our work can be extended to irregular surfaces, which shows a promising path for nanodiamond-based photonic sensors.
\end{abstract}
\begin{document}

\flushbottom
\maketitle
%
%
\thispagestyle{empty}
\setlength{\abovedisplayskip}{3pt}
\setlength{\belowdisplayskip}{3pt}
\section*{Introduction}

Mapping the low frequency (<1 kHz) magnetic fields and its vectorial orientation with micro and nanoscale resolution is crucial for many applications \cite{LeSage2013, Casola2018}. Various types of magnetic sensors have been developed on a variety of platforms in recent decades, such as superconducting quantum interference devices (SQUIDS) \cite{Denis_Vasyukov2013}, fluxgate devices \cite{Ripka}, and optically pumped magnetometers using alkali atom vapor cells \cite{Dehmelt, Bell&Bloom} that offer the measurement of magnetic field at high sensitivities. Generally, these sensors are limited in their spatial and sensitivity resolution due to their finite size, probing distance to the sample, and operating environment. In recent years, the negatively charged nitrogen-vacancy in diamond has emerged as one of the most promising tools for magnetic field sensing and imaging. The NV centers at room temperature offer high sensitivity for large NV ensembles with reported values that reach the $\mathrm{pT/\sqrt{Hz}}$ level \cite{Herbschleb2019, Bala2009} and high spatial resolution for single NV centers \cite{Taylor2008}. Numerous experimental studies with NV centers were carried out to detect electric fields \cite{dolde2011}, temperature \cite{acosta2010}, pressure \cite{Doherty2014}, and magnetic fields \cite{Sage2012, Maletinsky2012, Jensen2013} with spatial resolution down to the nanoscale. 

The spectacular advent of NV sensors is based on the very attractive optical and spin properties of NV color centers in diamond. The NV defect in a diamond lattice acts as a single isolated atom-like system consisting of a substitutional nitrogen atom with an adjacent vacancy. The electronic ground state of the NV center is a triplet state S = 1 with spin sublevels m$_{s}$ = 0 and m$_{s}$ = $\pm$1 \cite{Redman1991}. The spin sublevels m$_{s}$ = $\pm$1 are degenerate but are split from the spin state m$_{s}$ = 0 by roughly $D$ = 2.87 GHz \cite{Mason2006} at zero magnetic field and room temperature. The ground-level spin states of the NV center can be manipulated by microwaves and initialized and read optically due to state-dependent fluorescence \cite{Jelezko2004}. When an external magnetic field is applied, the degeneracy between the m$_{s} = \pm$ 1 states is lifted, and these states split proportionally to the applied field strength ($\sim$ 56 MHz/mT at low magnetic fields). Magnetic field detection with NV centers is performed by quantitatively measuring the Zeeman shifts \cite{Maze2008} using the optically detected magnetic resonance (ODMR) technique \cite{Wrachtrup1993}. Due to the C$_{3v}$ symmetry of the NV defect, the center quantization axis can take one of the four possible orientations in the diamond lattice. Depending on the orientation of the magnetic field relative to the four crystallographic NV axes, the ODMR signal from the NV ensembles will exhibit one to four pairs of resonance splitting between the states m$_{s}$ = 0 and m$_{s}$ = $\pm$1. When an applied magnetic field is oriented arbitrarily, each of the four NV crystallographic axes will sense a different field strength. In this scenario, the NV ensembles will exhibit four pairs of resonance splittings. Taking advantage of the position and angle dependence of the NV center quantization axis on the applied magnetic field and measuring the exact frequency of these resonance pairs, we can reconstruct the magnetic field vector in all spatial coordinates \cite{Steinert2010, Pham2011}.

To map magnetic fields at high sensitivity and high spatial resolution, the diamond NV sensor should be in close proximity to the sample (within a few 10 nm). For that reason, conventional magnetometry measurements are done typically with NV 
 \begin{figure}[H]
\centering
\subfloat[]{\label{}\includegraphics[scale=0.65]{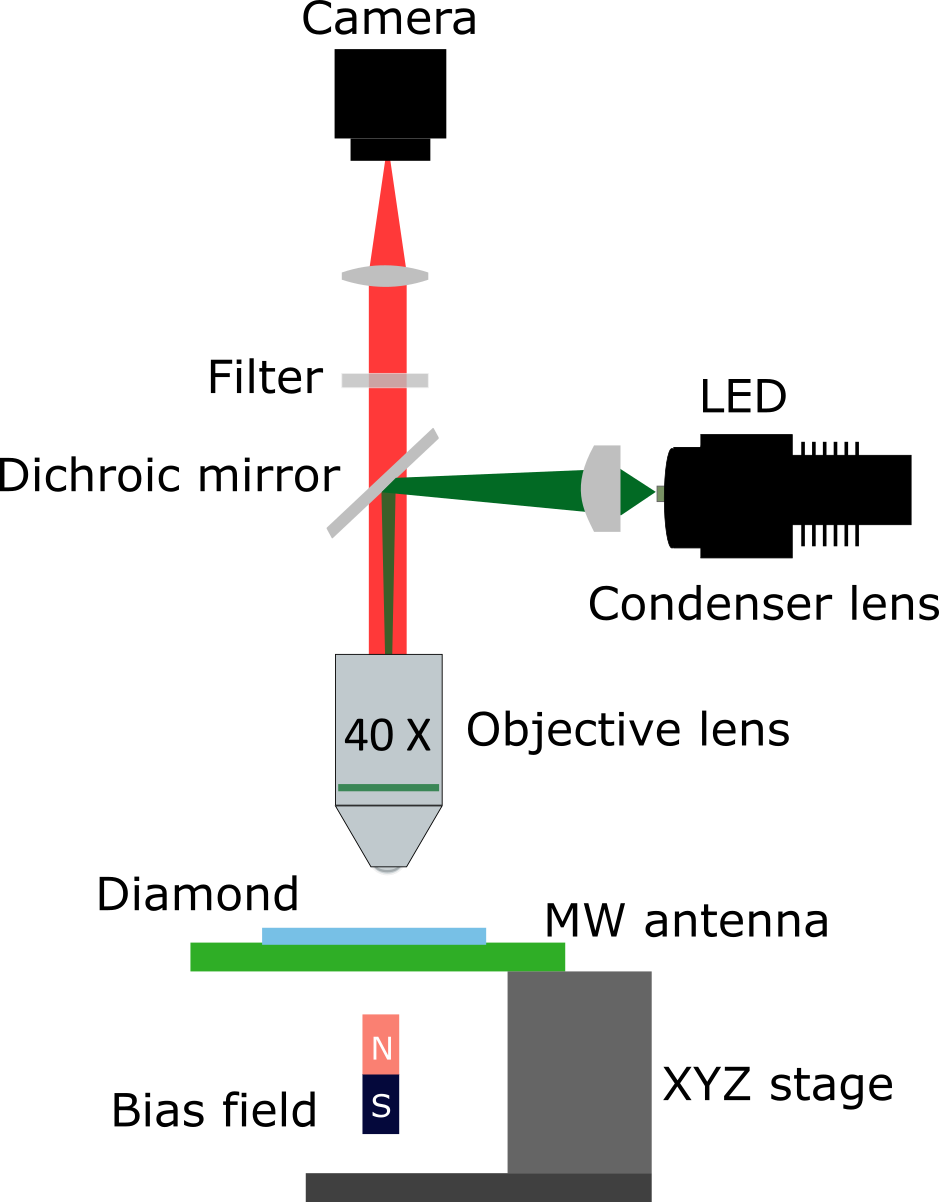}}\quad
\subfloat[]{\label{}\includegraphics[scale=0.45]{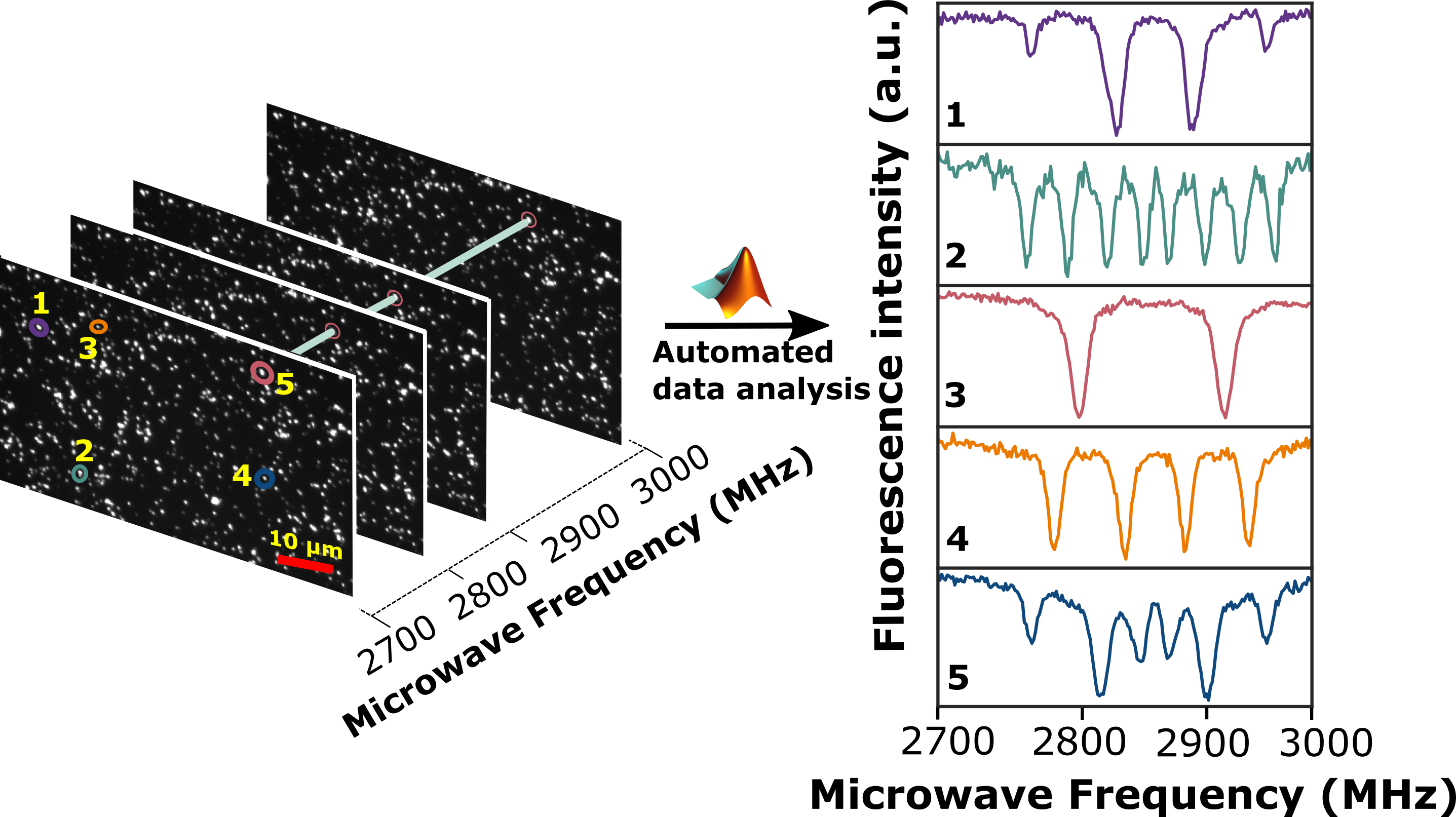}}
\caption{(a) Schematic of the wide-field diamond magnetometer used for the described magnetic field sensing and imaging. (b) Layout of the volumetric image data from the ODMR imaging measurement, where a total of N frames are obtained by sweeping the microwave frequency around 2.87 GHz with rows and columns of pixels that form 2D slices with N frames that form the volume. The ODMR signal is obtained by averaging $n$ pixels (indicated in blue) in the pixel region. The original volumetric data contain the ODMR spectrum for each diamond spot. Plot of ODMR spectra extracted from the image data using the automated algorithm for five different diamond spots (numbered 1 to 5). Each of the diamond crystals is oriented at different angles with respect to the magnetic field direction.}
\label{fig:scheme}
\end{figure}
ensembles in a bulk diamond with a thin NV layer, or a scanning atomic force microscopy (AFM) probe in the form of a diamond tip is used, which requires a smooth sample surface and proximity of the probing device, often limiting the sensing capabilities. Our approach in this experimental work is to use randomly oriented nanodiamonds (NDs) with NV color centers to realize a simple and inexpensive wide-field magnetometer. Because of their small size, nanodiamonds can be adhered to any irregular material surface, even including fiber tips and living cells. The nanodiamonds offer a trade-off between sensitivity, resolution, and manufacturing costs. An additional important advantage of our approach is the use of wide-field magnetic imaging, which offers in-time parallel mapping of the magnetic field over a large field of view and a shorter measurement time. In this paper, we address the feasibility of utilizing NV centers in nanodiamonds to map the true three-dimensional orientation of magnetic fields. 

 Moreover, we show the magnetometer's applicability by spatially mapping the magnetic field distribution of a DC field from a straight thin current-carrying wire placed in near proximity to the diamond sample. Our approach to estimating the magnetic field is an extension of the work reported in \cite{Chipaux2015}, where a narrow NV diamond slab has been used to measure the magnetic field. In this work and as a proof-of-concept experiment, we probed the NV centers in randomly oriented 1 $\mu$m-sized microdiamonds that were deposited on a planar glass substrate using the matrix-assisted pulsed laser evaporation technique (MAPLE). The distinctive advantage of the MAPLE technique compared to other deposition techniques (such as spin coating, dip coating, sequential assembly, etc.) is the ability to obtain thin films of uniformly distributed nanodiamonds on both planar and nonplanar surfaces. We measured the ODMR signals in such randomly oriented diamonds and used them to perform scalar and vector magnetometry relative to the diamond lattice. In general, this method can be extended to nanodiamonds deposited on any material surface. The deposition of nanodiamonds on any material surface offers a low-cost solution with fewer experimental complexities, and the results obtained in this work show a promising path for novel and inexpensive nanodiamond-based photonic sensors for a multitude of material and biological applications.

\section*{Results and discussion}
\subsection*{Wide-field ODMR magnetic imaging}
Figure \ref{fig:scheme}a describes the wide-field diamond magnetometer setup used to perform the ODMR magnetic imaging of NV centers in randomly oriented nanodiamonds. The magnetic imaging scheme works as follows; the excitation of ground-state spins is achieved by continuously illuminating the diamond sample with a 70 mW pump beam of 530 nm. Under continuous light illumination, the microwave frequency is swept around the $m_{s} = \pm 1$ transition resonance frequency of 2.87 GHz. The fluorescence (600 - 800 nm) signal emitted from the NV centers was then recorded by taking a single snapshot of the sample field of view (FOV) for each MW frequency. This results in volumetric image data with rows and columns of pixels forming 2D slices with N frames that form the volume. The ODMR signals were simultaneously imaged from all fluorescent diamond spots over the entire FOV of the objective in one exposure shot. Such wide-field imaging is faster and more efficient than confocal scanning using an NV probe on an AFM \cite{Ficek2021}, which typically requires hours of measurement time to scan the entire sample area.
 \begin{figure}[H]
\centering
\includegraphics[scale=0.40]{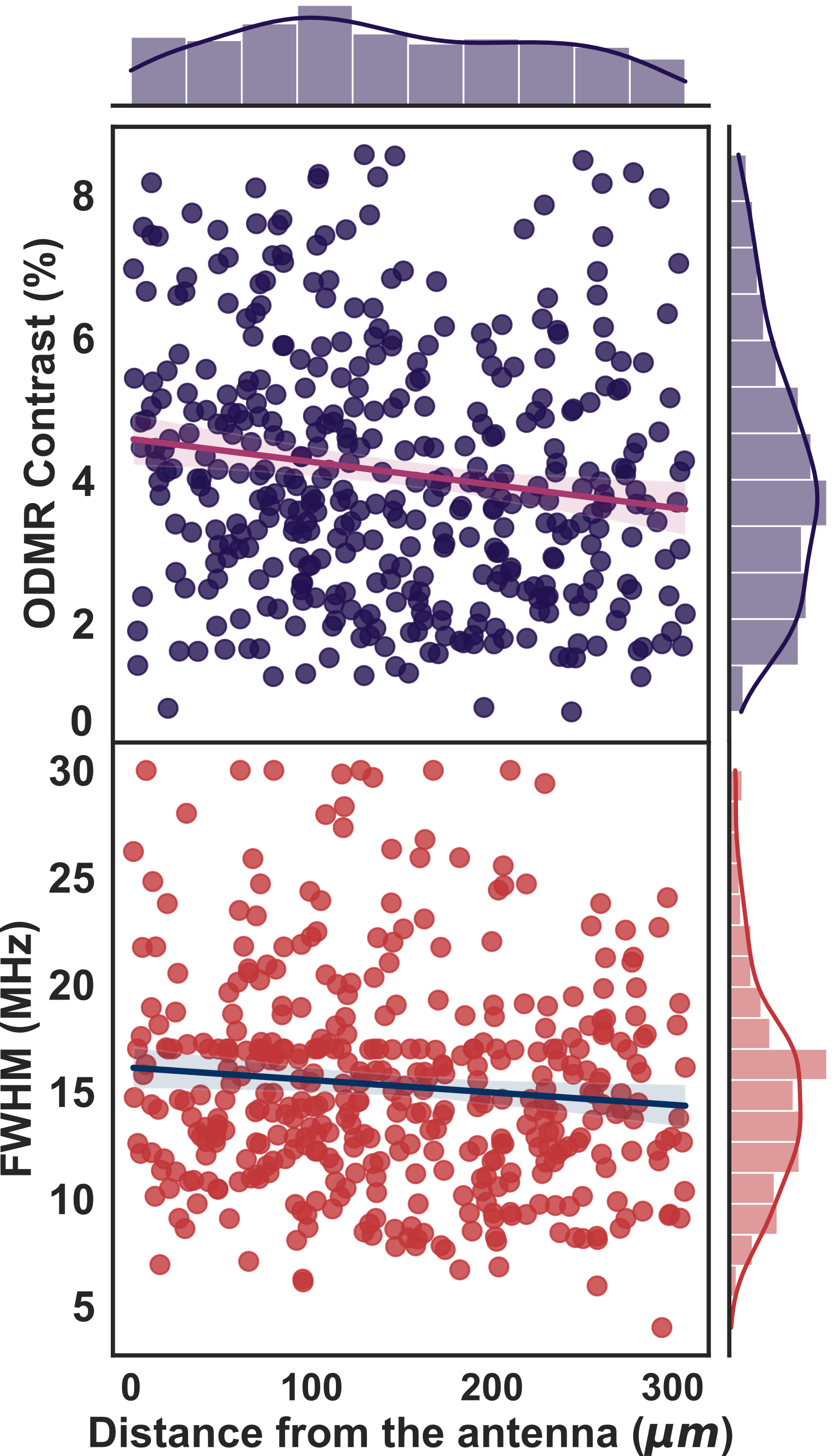}
\caption{ODMR contrast (top panel), and FWHM (bottom panel) of all diamond spots as a function of the distance from the MW antenna. The side panels show the histogram plotted with the kernel density estimation.}
\label{fig:kde}
\end{figure}
\subsection*{Automated multipixel extraction of the ODMR spectrum}

To recover ODMR signals from volumetric image data, we employ an internally developed automated MATLAB \cite{MATLAB} algorithm. Bright fluorescent diamond spots were first identified from a single image of the data. Each identified diamond spot is then characterized by its spatial position (XY), spot area (pixel region), and orientation in the XY plane (see supplementary Fig. S1 online). In addition, spots are sequentially numbered for future identification and association with the corresponding ODMR spectrum. The pixel region is predetermined by the algorithm when identifying the diamond spots, and the size of the pixel region varies depending on whether a single microdiamond or an aggregate is observed. The ODMR signals are then obtained in the following manner: for a given diamond spot, the ODMR spectrum is extracted by averaging the pixel values within the pixel region. For a single diamond spot, the mean pixel intensity obtained from a single image gives a single data point on the ODMR spectrum. Since each image corresponds to a certain microwave frequency, the procedure is then repeated for all images, resulting in a complete ODMR spectrum. 

 The data analysis procedure is fully automated from the initial reading of the volumetric image data, identifying the diamond spots, retrieval of ODMR spectra, fitting the resonance peaks, to finally estimate the magnetic field intensities, and does not require visual analysis or human intervention. An example of extracted ODMR spectra using the automated algorithm in the presence of a small bias field (around 3 mT) is shown in Fig. \ref{fig:scheme}b for five randomly oriented diamonds. Each of the diamond spots exhibits a different number of splittings as a result of random crystallographic orientation with respect to the magnetic field direction. The zero-field ODMR contrast and the full width at half maximum (FWHM) (on each axis side showing the kernel density estimation overlaid with marginal histograms to visualize the distribution of each variable) for all diamond spots in the field of view is shown in Fig. \ref{fig:kde}. We observed a variation in the contrast of the ODMR signal between 1\% and 8\% with a mean value of 4.5 \%. The variation is due to the dependence of ODMR contrast on the number of NV centers in a diamond spot, the size of the diamond aggregate, and the microwave power at the location of the diamond spot. 
 
 In our measurement, we observed about 410 diamond spots over a field-of-view area of 307 x 245 $\mathrm{\mu m^{2}}$ corresponding to 1280 by 1024 pixels. Each pixel in our wide-field setup corresponds to an area of (0.24 $\mathrm{\mu m)^{2}}$. Given the mean size of the micro-diamonds is 0.75 $\mathrm\mu$m, and there is additional blur due to optical diffraction ($\sim 0.6~\mathrm{\mu}$m), we estimate that a single micro-diamond image on the sensor is 4-5 pixels wide. However, some diamonds are out of focus and hence appear much larger on the camera sensor. While the overall number of diamonds in the field of view could be further increased, this will result also in the rejection of more overlapping diamond cases.

The average FWHM value of the linear fit is estimated to be 16 MHz. The larger line width is caused by a high microwave power that causes power broadening of the ODMR spectrum.
\begin{figure}[H]
    \centering
    \subfloat[]{\label{}\includegraphics[scale=0.2]{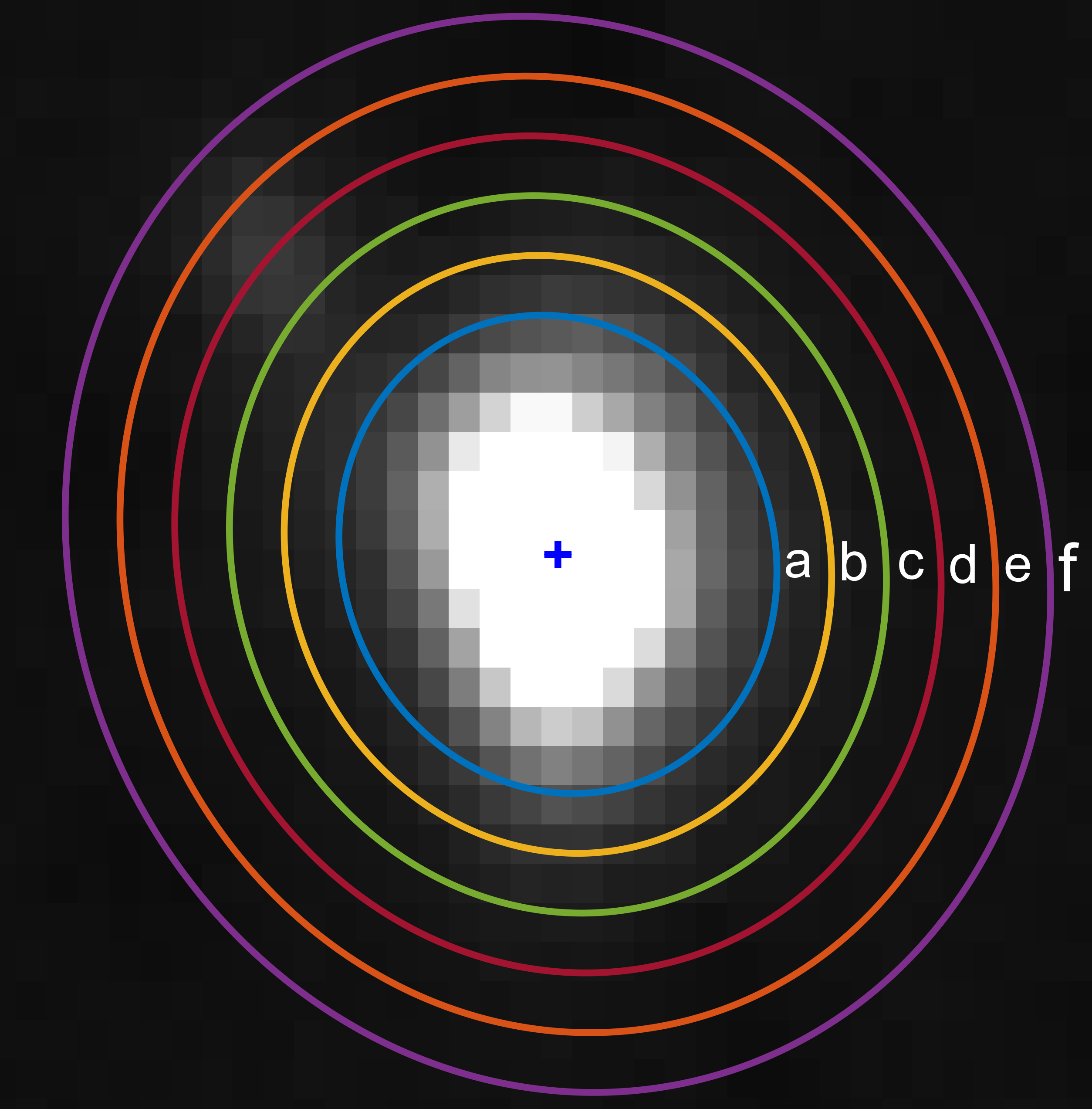}}\qquad
    \subfloat[]{\label{}\includegraphics[scale=0.5]{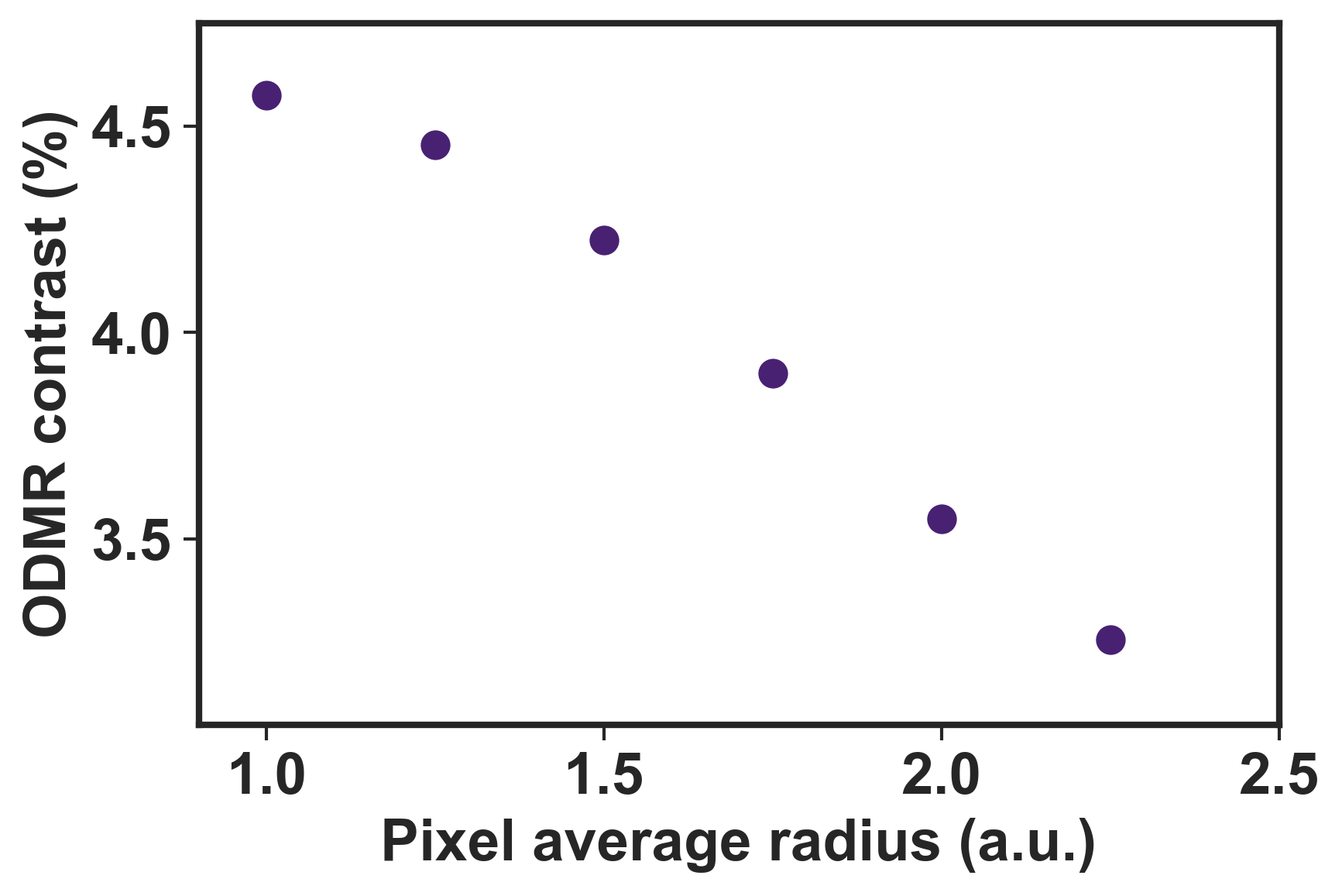}}\\
    \subfloat[]{\label{}\includegraphics[scale=0.5]{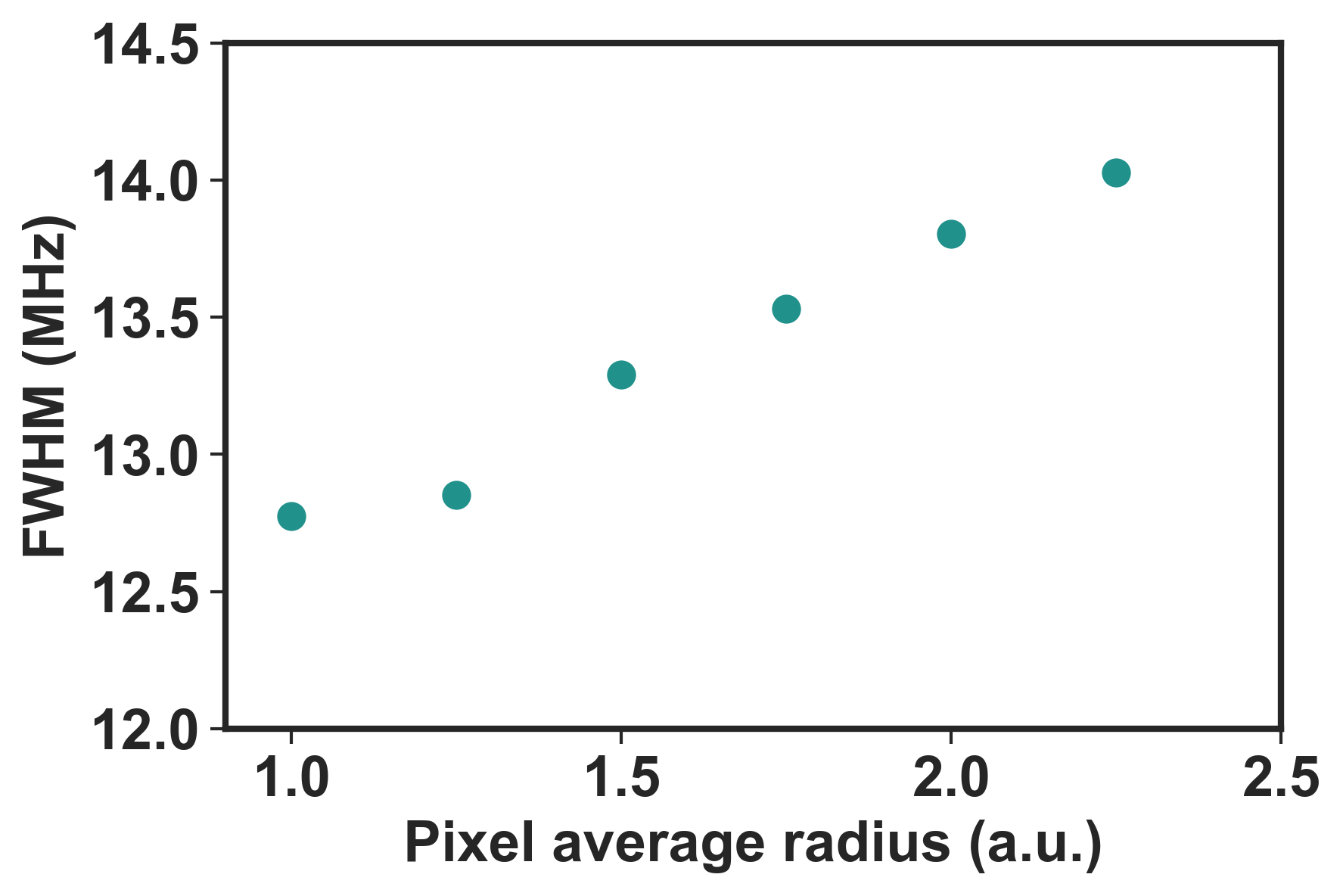}}\quad
    \subfloat[]{\label{}\includegraphics[scale=0.5]{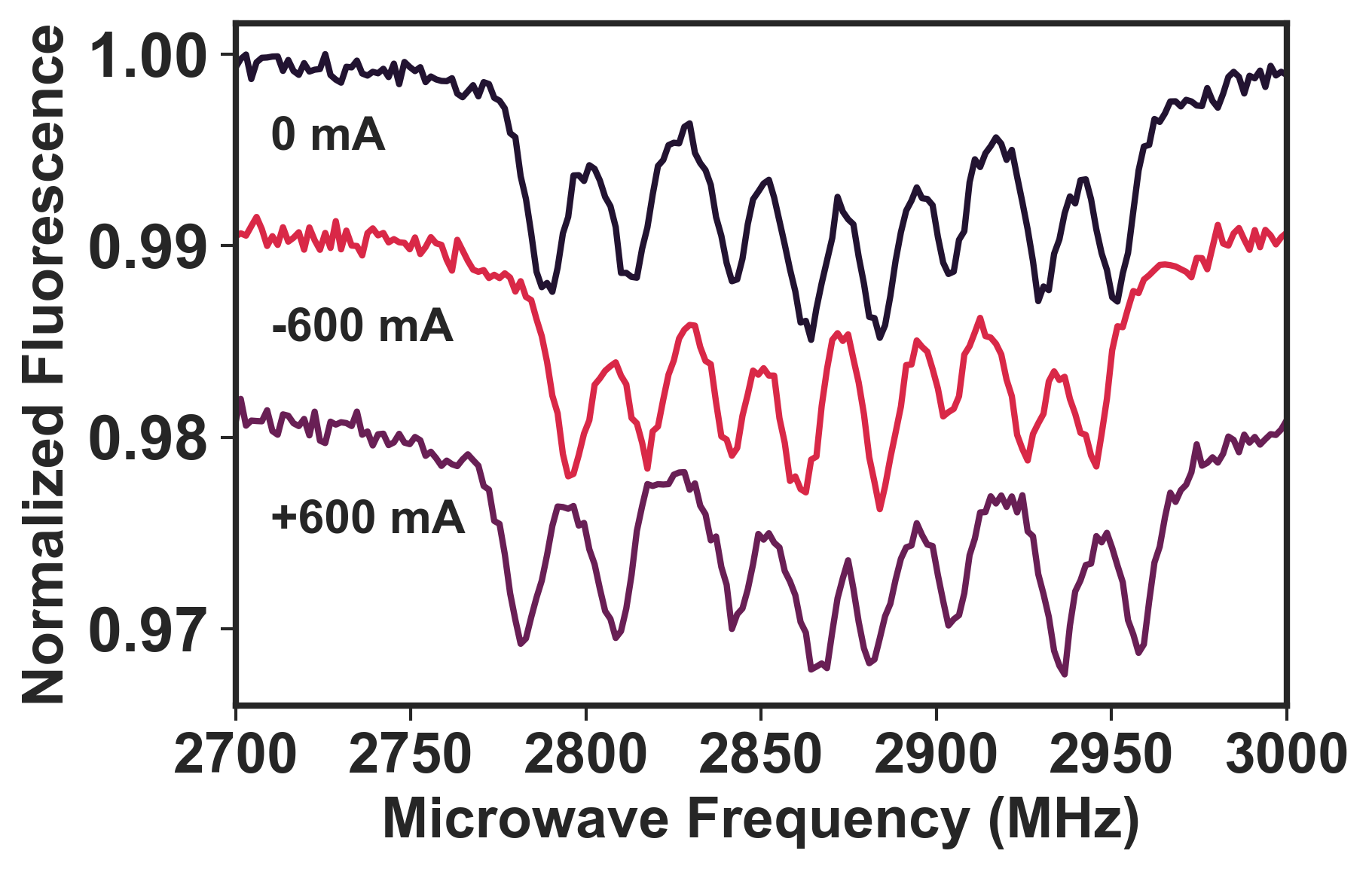}}
    \caption{(a) A fluorescent diamond spot with different pixel region radius encircled in different colors and the corresponding ODMR contrast (b) and FWHM (c) for different pixel region. The pixel areas named from a to f in panel a corresponds to the pixel region radius from 1 to 2.5 in panels b and c. (d) ODMR resonance spectra in presence of a bias field at 0 mA (black) and 600 mA (red) and -600 mA (violet) in the current-carrying wire.}
    \label{fig:roi}
\end{figure} 
\subsection*{Effect of pixel region and positional drift}

An important consequence of the automated data analysis procedure is the effect of incorrect pixel region and positional drift of diamond spots on the extracted ODMR signal. Both of these effects can adversely affect the ODMR contrast and linewidth of the extracted ODMR spectrum from the image data, which leads to inaccurate estimation of the magnetic field values. Therefore, it is necessary to validate the predetermined pixel region by the algorithm so that the extracted ODMR spectrum has maximum contrast and minimal FWHM. In this study, we performed an analysis on several diamond spots with various pixel regions (see Fig. \ref{fig:roi}a) and found that for most of the diamond spots, the pixel region estimated by the algorithm is accurate enough to obtain the maximum ODMR contrast and the minimal line width as shown in Fig. \ref{fig:roi}b and Fig. \ref{fig:roi}c respectively. Often, during the analysis of the image data, we observed a drift in the XY position of diamond spots, typically by a pixel or two, between successive images and repeated scans. This positional drift is attributed to the thermal instability of the glass substrate that hosts the diamond crystals and is caused by radiation-induced heat from the microwave antenna and resistance-induced heat from the current wire, especially at higher current intensities during the wide-field magnetic imaging measurements. In addition, the continuous illumination of the laser adds up to sample heating as well. Positional drifts between frames are corrected by properly identifying the same fluorescent spot in successive images by finding the minimum distance d $d(p,q)^{2} = (x_{q}^{2} - x_{p}^{2}) + (y_{q}^{2} - y_{p}^{2})$ between two diamond spots p and q and thereby updating the XY coordinates for each image with a constant pixel region while averaging the pixels values for all images. With this distance minimization approach, we eliminate the typical problems associated with image translation and rotation.

\subsection*{Estimation of magnetic field from a current-carrying wire}
\begin{figure}[H]
    \centering
    \subfloat[]{\label{}\includegraphics[scale=0.45]{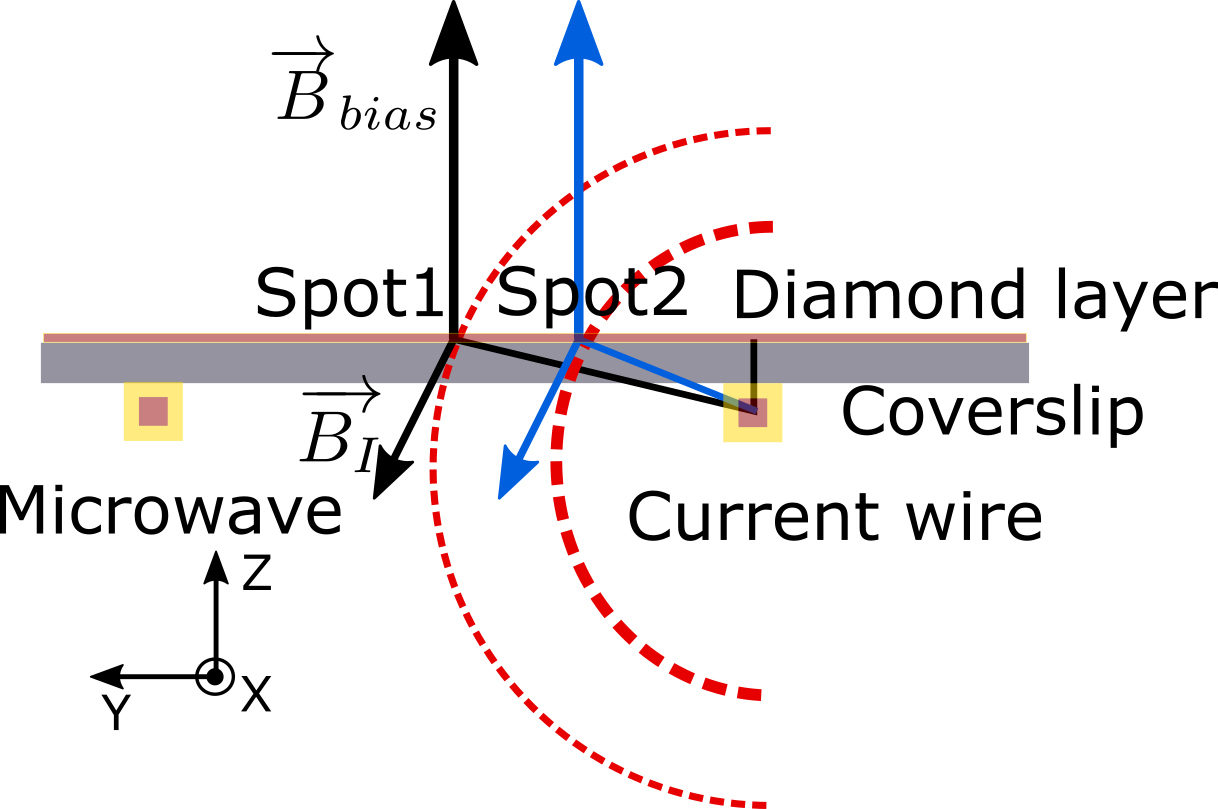}}
    \subfloat[]{\label{}\includegraphics[scale=0.45]{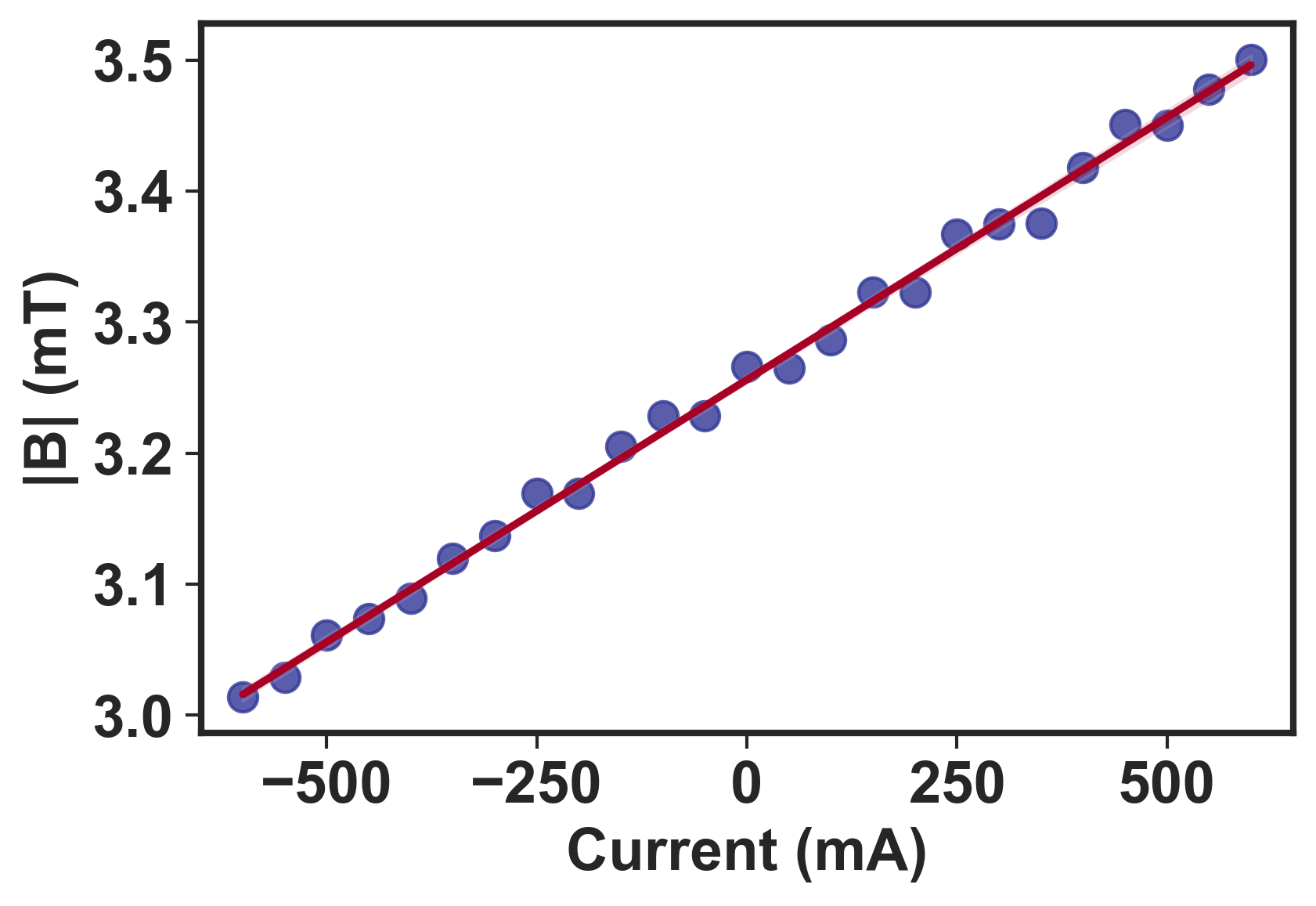}}
    \subfloat[]{\label{}\includegraphics[scale=0.45]{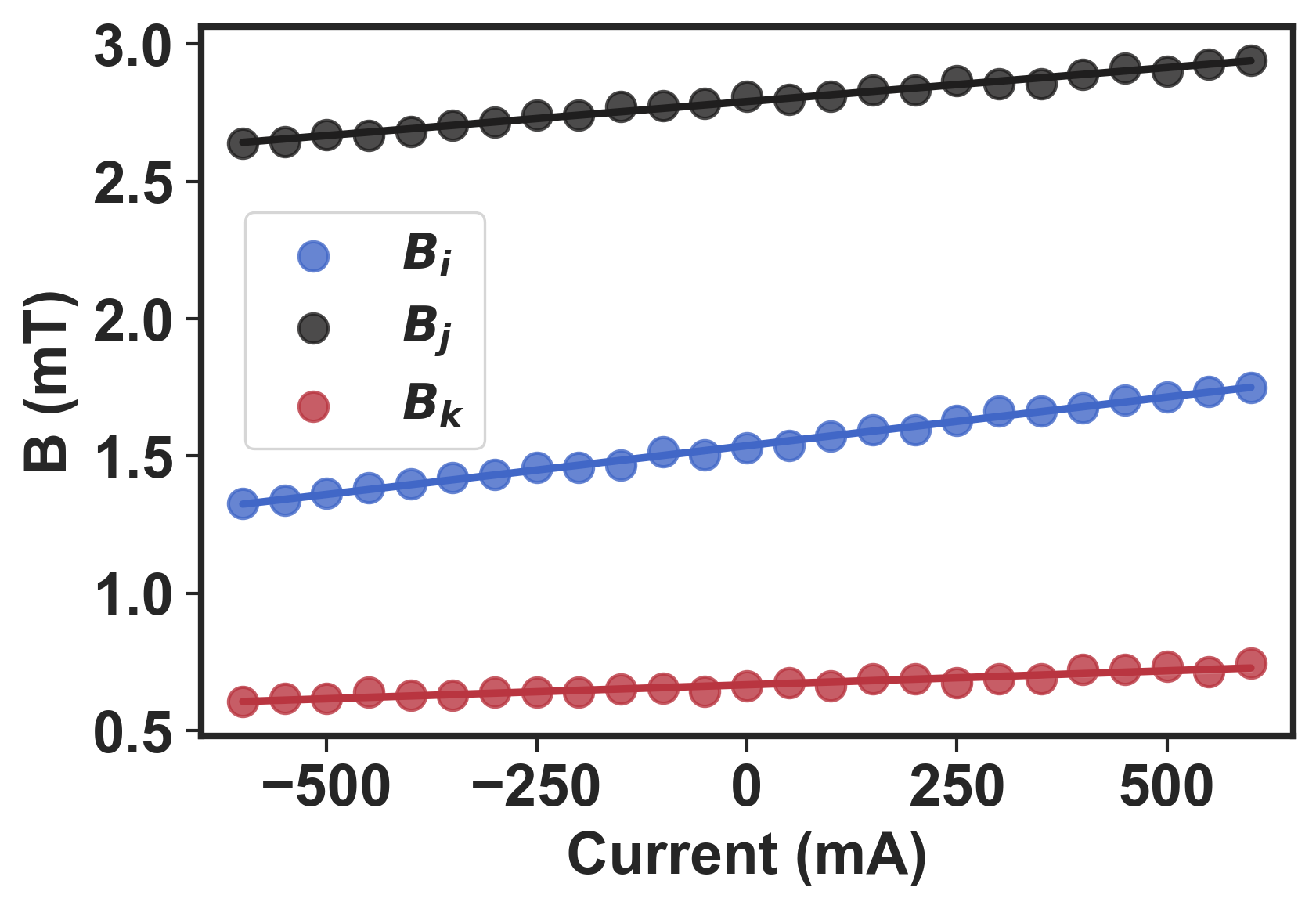}}\\
    \subfloat[]{\label{}\includegraphics[scale=0.40]{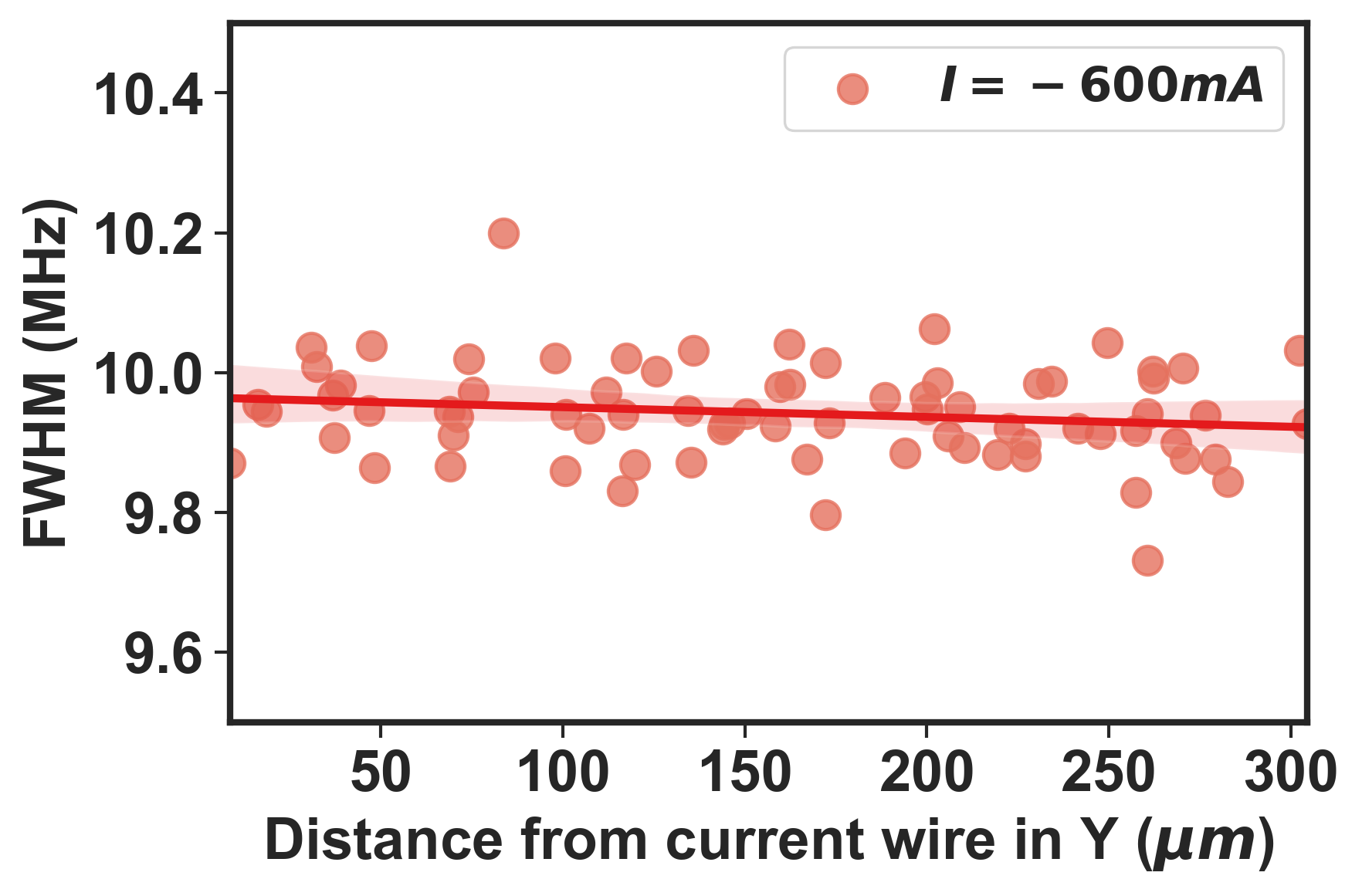}}
    \subfloat[]{\label{}\includegraphics[scale=0.40]{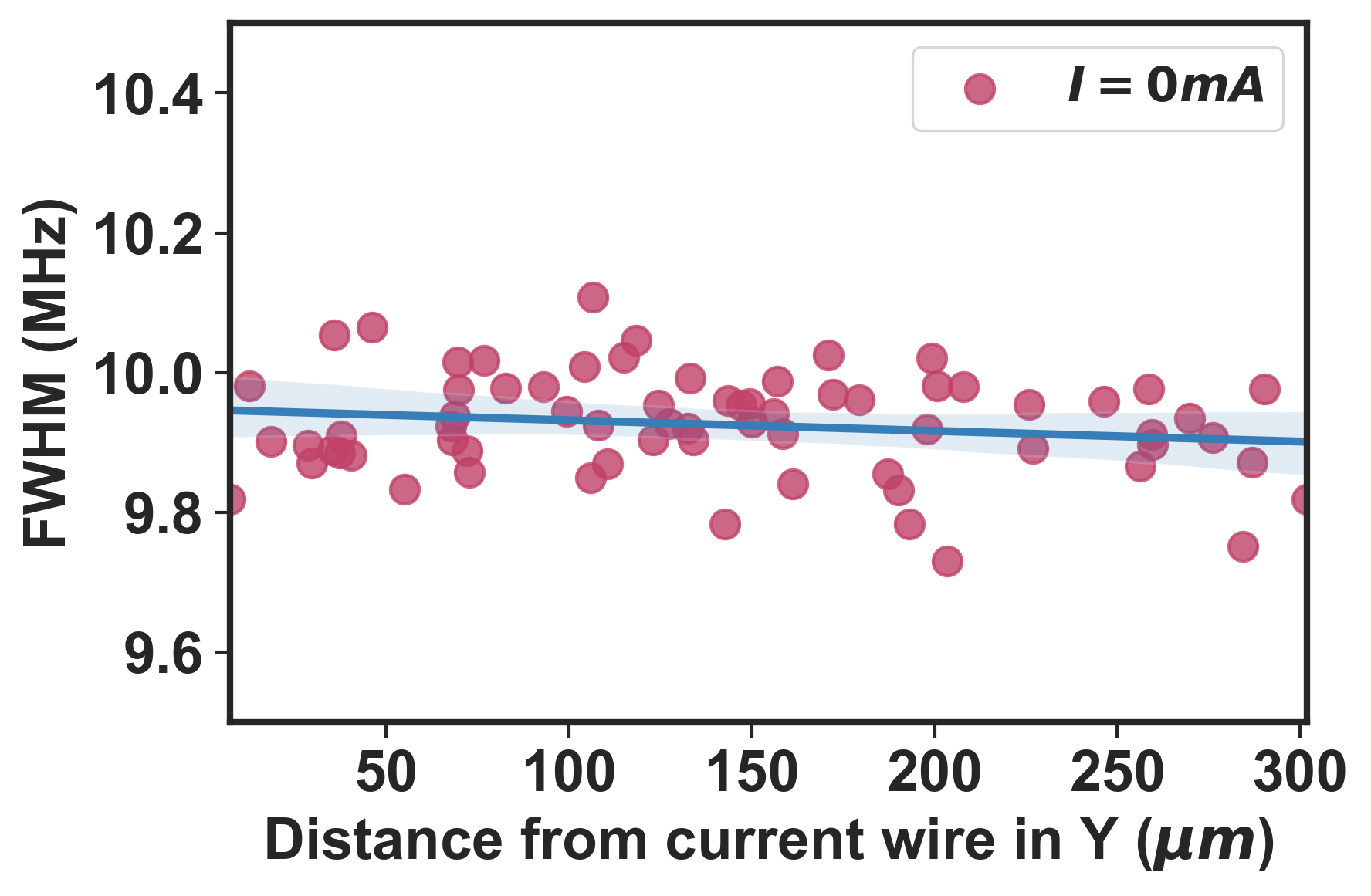}}
    \subfloat[]{\label{}\includegraphics[scale=0.40]{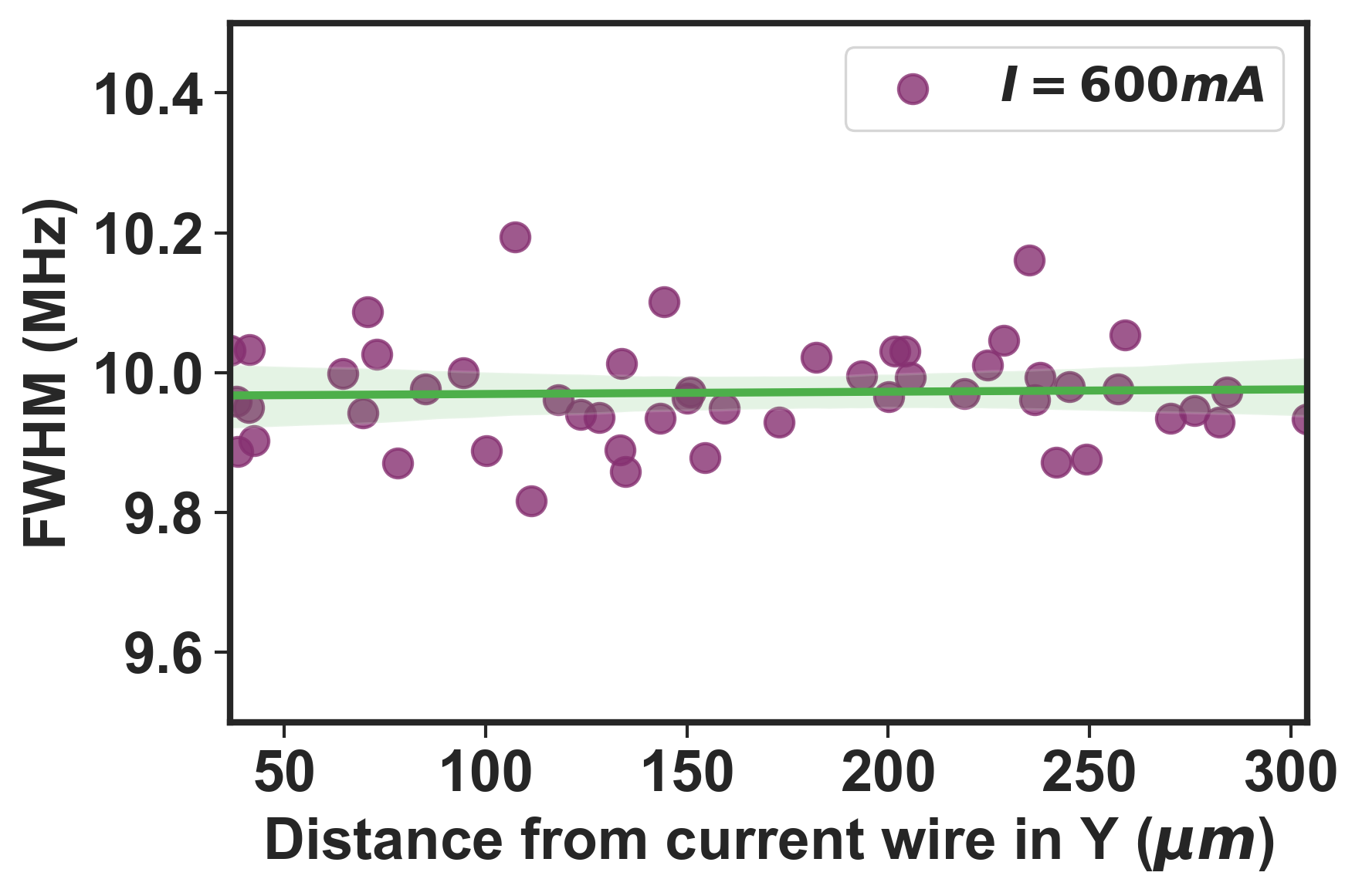}}\\
    \subfloat[]{\label{}\includegraphics[scale=0.45]{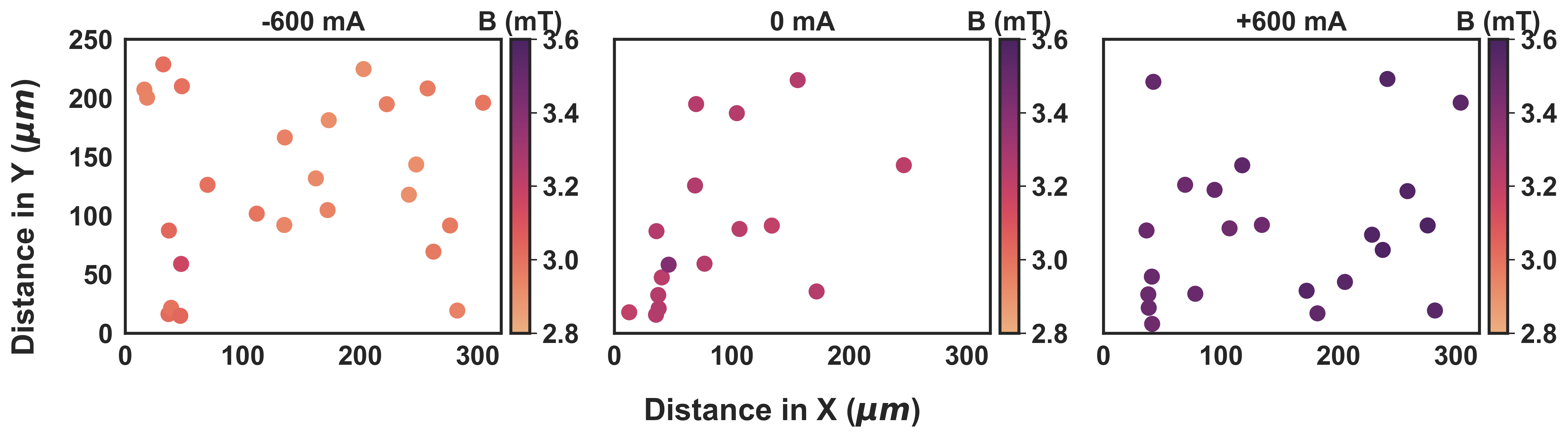}}
    \caption{(a) Scheme of the arrangement of thin diamond layer (red), coverslip (grey), microwave and current wire (orange) with magnetic field lines (dashed magenta). $B_{I}$ and $B_{bias}$ denote the direction of the magnetic field vector for bias and current wire, respectively. (b) The three individual field components measured from a single diamond spot. (c) The total magnetic field value is plotted against different current intensities for a single diamond spot. (d-f) FWHM extracted from the Lorentzian fit for arbitrarily oriented diamond spots for current values -600 mA, 0 mA and 600 mA respectively. (g) Map of magnetic field strength for current set at -600 mA, 0 and 600 mA with the field strength indicated in the colorbar. The scatter points indicate the XY position of the randomly oriented micro-diamond crystal in the field-of-view. The current wire is situated to the right of the field map.}
    \label{fig:bmaps}
\end{figure}
As a result of the random orientation of the diamond crystals, the four crystallographic axes of the NVs are also randomly oriented with respect to the applied bias field direction. This implies that for a given set of N diamond spots, it is observed that a certain population has four discernible pairs of ODMR resonances, while some populations exhibit partially degenerate ODMR resonances associated with specific crystallographic directions such as \{111\}, \{100\} and \{110\} families. The extracted ODMR spectrum is fitted to a sum of eight Lorentzians equation (\ref{eqn:LF}) using the Levenberg-Marquardt (LM) \cite{More_JJ} algorithm to obtain the exact resonance frequencies (see supplementary Fig. S2 online). The LM algorithm requires initial guesses for the fitting function to work that are not too arbitrary from the final result. The initial fit parameters are predetermined fully by the automated script taking advantage of the MATLAB inbuilt functions. For estimating the magnetic field strength, only the diamond spots that exhibit four pairs of resonances were taken into account. 
\begin{equation}
\centering
    f(v) = 1-\sum_{i=1...8}C_{i}\left(\frac{\gamma^{2}}{4(x_{i}-x_{c})^{2}+\gamma^2}\right)
\label{eqn:LF}
\end{equation}
where C$_{i}$ is the contrast, $\gamma$ is the full width at half-maximum (equal for all resonances), and $x_{c}$ is the central position of a single resonance. 
The precise parameters obtained from the fit were used to determine the bias magnetic field strength and the field generated by the current wire. The $^{14}$ N hyperfine splitting could not be resolved in our measurement using our wide-field setup. Moreover, the four pairs of resonances were not well resolved because of the limitation in the generation of higher radio frequencies of our MW generator, although the magnetic resonance peaks are clearly visible with a few MHz separation between each peak.

Following the Ref. \cite{Chipaux2015} for calculating the magnetic field strength, the diamond lattice coordinate frame (i,j,k) is chosen such that the unit vectors u$_{a}$, u$_{b}$, u$_{c}$, u$_{d}$ have the following coordinates: $u_{a}=\frac{1}{\sqrt{3}}\left(1,1,1\right)$, $u_{b}=\frac{1}{\sqrt{3}}\left(-1,-1,1\right)$, $u_{c}=\frac{1}{\sqrt{3}}\left(-1,1,-1\right)$, $u_{d}=\frac{1}{\sqrt{3}}\left(1,-1,-1\right)$. The magnetic field components B$_{i}$, B$_{j}$, B$_{k}$ to be measured are given by the magnetic field projection m$_{a}$, m$_{b}$, m$_{c}$, m$_{d}$ on the four NV quantization axes, respectively. Thus, we obtain the following.  
\begin{equation}
\centering
\begin{split}
        B_{i} &= \frac{\sqrt{3}}{4}(-m_{a}+m_{b}+m_{c}+m_{d})\\
        B_{j} &= \frac{\sqrt{3}}{4}(-m_{a}+m_{b}+m_{c}+m_{d})\\
        B_{k} &= \frac{\sqrt{3}}{4}(-m_{a}+m_{b}+m_{c}+m_{d}).
\end{split}
\label{eqn:vector}
\end{equation}
Finally, the norm of the magnetic field vector $\overrightarrow{B}$ gives the total magnetic field strength sensed by the NV ensembles, given as $\left|B\right| = \sqrt{B_{i}^{2} + B_{j}^{2} + B_{k}^{2}}$.

In the first set of experiments, with properly optimized experimental settings (laser power output, bias field, and MW frequency bandwidth), we performed the ODMR magnetic imaging in the presence of a bias field and zero current. We observed a linear dependence of resonance splitting as a function of the magnetic field, and the bias field strength was determined using equation (\ref{eqn:vector}). The estimated value of the bias field from the measurement is 3.2 mT. Once the bias field is estimated, we repeat the ODMR measurement for different current intensities on the same FOV of the diamond sample. A DC current in the range of -600 mA to 600 mA is applied to one of the dual stripline of the MW structure, which was placed directly below the diamond sample such that the current wire lies in the x direction. The MW structure is oriented so that each of the strip lines occupies the start and end of the objective FOV. The NV centers senses both the magnetic field generated by the current wire and the applied bias field from the permanent magnet, and the total B-field is given by $\overrightarrow{B}_{NV} = \overrightarrow{B}_{bias} + \overrightarrow{B}_{I}$. Figure \ref{fig:roi}d show the ODMR plots obtained from a single diamond spot for current values -600 mA and 600 mA together with the zero-current (0mA) ODMR spectrum. It is evident from the plots that the total magnetic field intensity sensed by NV ensembles is slightly higher for 600 mA and lower when the direction of the current is reversed for the same current value. 

The current wire produces an ortho-radial magnetic field in the yz plane. Depending on the direction of the current flow in the wire, $\overrightarrow{B}_{I}$ adds or subtracts from the field $\overrightarrow{B}_{bias}$. 
The diamond sensor with the NV centers was deposited on a glass cover slip with a thickness of 1.7 mm, which introduces a small height in the z direction, as shown in the schematic Fig. \ref{fig:bmaps}a. The magnetic field sensed by the NV no longer decreases by $1/r$ from the center of the current wire. The additional small height in the z-direction introduces an angle dependence on the resultant magnetic field vector. The diamond spots (spot1 and spot2) sense different magnetic field strengths depending on their position in the XY plane. Due to the non-circular geometry of the current wire, the angle dependence of $\overrightarrow{B}_{I}$ could not be accurately estimated. The total magnetic field and components of the magnetic field in the NV diamond frame for a single diamond spot are shown in Fig. \ref{fig:bmaps}b and Fig. \ref{fig:bmaps}c, respectively. It is clearly seen that the magnetic field sensed by the NV centers increases with increasing current intensity and decreases when the current is reversed. Additionally, we measured the FWHM for all diamond spots shown in Figs. \ref{fig:bmaps}d-\ref{fig:bmaps}f that were arbitrarily oriented to the magnetic field and found its value to be 10 $\pm$ 0.3 MHz. No significant broadening of the ODMR peaks was observed even at higher currents, and the FWHM of the diamond spots at 600 mA was consistent with the FWHM at 0 mA. This suggests that our arbitrarily oriented diamond sensor can work in a wide range, allowing us to detect smaller field fluctuations. The spatial magnetic field distributions estimated using NV centers in multiple diamond spots for three different current values -600 mA, 0 mA and 600 mA are shown in Fig. \ref{fig:bmaps}g. 
\begin{figure}[H]
    \centering
    \includegraphics[scale=0.50]{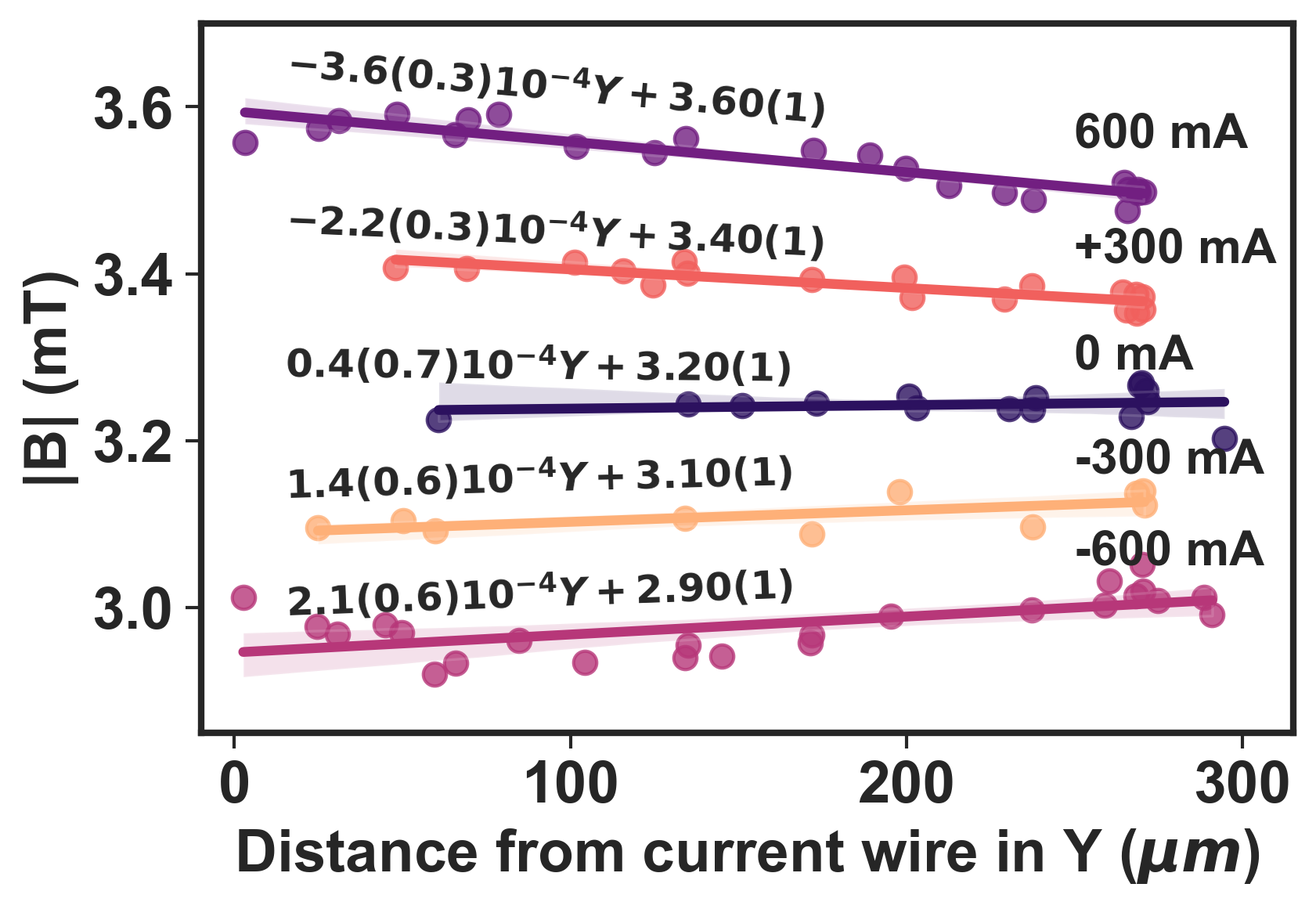}
    \caption{The magnetic field strength plotted as a function of distance from the current wire in Y direction for different curve values. The colored solid lines are the linear-fit to the data and the shaded area show the 95 \% confidence bounds. }
    \label{fig:bcurr}
\end{figure}
Figure \ref{fig:bcurr} shows the total magnetic field strength estimated from the ODMR spectrum of diamond NV crystals that are spatially scattered across the FOV for different current intensities. The B-field values are plotted as a function of the distance from the current-carrying wire. As expected, and it is evident from the estimation that the higher the current value, the greater the magnetic field. The bias field with zero current (0 mA) in the wire is shown as a reference. A small total magnetic field gradient of 0.1 mT/300 $\mu$m is observed in the y direction. However, no gradient was observed for lower current values. Fluctuations in the estimated values of the total magnetic field strength are attributed to the magnetic noise induced in the measured environment. The effect is more pronounced at higher currents, probably caused by the crosstalk between the MW strip lines or by the heating of the wire due to the high DC current.

The magnetic field sensitivity of a wide-field cw-ODMR imaging magnetometer is given by \cite{JFBarry2020, Dreau2011} 
\begin{equation}
    \eta_{cw} \approx P\frac{h}{g_{e}{\mu_{B}}}\frac{\Delta\nu}{C_{cw}\sqrt{R}}
\end{equation}\\
with $\Delta\nu$ being the ODMR line width, R the photon-count rate, $C_{cw}$ the ODMR contrast and the prefactor P depends on the specific resonance line shape which is 4/(3$\sqrt{3}$) when assuming a Lorentzian line shape. Experimentally, the sensitivity is attributed to the ratio of contrast and the linewidth $C/\Delta\nu$, i.e., the steepest slope of the ODMR resonance line shape. Based on the measured data, the estimated sensitivity is 4.5 $\mathrm{\mu}T/\mathrm{\sqrt{Hz}}$ and is severely limited by the low light intensity of LED source used in the wide-field setup, the limited pixel well capacity and relatively low signal-to-noise achievable on the camera sensor, and only a moderate numerical aperture of the objective used for fluorescence light collection.

\section*{Conclusion}

In this paper, we developed a widefield diamond magnetometer using NV centers in randomly oriented 1 $\mu m$ diamond powder capable of simultaneously measuring the spatially varying DC magnetic fields over large fields of view ($\sim$ 300~$\mu$m). We imaged the ODMR signals and reconstructed the vector magnetic fields relative to the diamond reference frame. Additionally, we have developed an automated MATLAB algorithm for extracting the ODMR spectrum and estimating the magnetic field for a large number of bright diamond spots from the wide-field images. Moreover, as a general demonstration, we employed this NV sensor to measure the magnetic field generated by a current-carrying wire. Our experimental results show that the measured magnetic field value with an NV sensor for a given current intensity agrees well with the theoretically estimated value. In summary, our experimental results show a promising path for vector magnetometry using arbitrarily oriented diamond crystals. Our approach to use randomly oriented nanodiamonds could pave the way for low-cost, rapid mapping of magnetic fields of interest on photonic platforms, which may find applications in biomedical imaging. 

However, we note there are certain limitations in achieving a high-resolution map of the B-field distribution using the presented methodology. The first constraint is due to the limited fraction of microdiamonds that are oriented arbitrarily to the applied magnetic field, i.e. in such a way that none of the lines overlap in the ODMR spectrum, which limits the spatial mapping of B-field to a sampling of  randomly located points. At the same time, the diamond particle density has to be kept below a level of cluster formation in order to be able to optically address individual particles. On the other hand, significant improvements can be made to the current experimental setup to enhance the sensitivity, e.g. by using pulsed ODMR techniques and by increasing the light collection efficiency using a higher NA objective. 

Looking forward, an interesting possibility is to implement an absolute vector magnetometry using these arbitrarily oriented diamond powders. This can be achieved by precalibrating each micro-/nano-diamond orientation in the laboratory frame with magnetic fields applied in several specific directions, e.g. using a 3D Helmholtz coils set. With a known diamond orientation, the methodology presented here enables one to determine the full B-field vector. This extension could open up even more applications in different areas of research.  

\section*{Methods}

\subsection*{The diamond sample}
The main criteria for selecting a suitable nanodiamond sample for magnetic sensing are the sample volume, high concentration of NV centers, proximity of NV centers to the surface, and the uniform distribution of nanodiamonds on the surface of the material. Great effort has been made to obtain thin films of uniformly distributed nanodiamonds using various deposition techniques (spin coating, dip coating, sequential assembly, etc.).  Such a diamond sample could be conveniently prepared by the laser ablation process in diamond micro-or nanoparticles, the so-called Matrix-Assisted Pulsed Laser Evaporation (MAPLE) \cite{Sawczak} deposition technique. The MAPLE method is a derivative of pulsed laser deposition (PLD), but instead of direct material ablation, a cryogenically frozen suspension of the deposited nanocrystalline material fulfills the role of the target being transferred onto suitable substrates as a result of pulsed laser-induced evaporation. 

The base target for laser ablation in our experiments was prepared by suspending fluorescent microdiamond powder (MDNV1umHi30mg, Adamas Nanotechnologies) in deionized water. The suspension was solidified at cryogenic temperature (90-100 K) and served as a target in the MAPLE process. The deposition was carried out under vacuum conditions ( $10^{-5}$~mbar) with a 3225 nm laser. Thin films were made on glass coverslips (1 × 1 cm$^{2}$), mounted at a distance of 15 mm from the target surface and preheated at temperatures in the range of 293-323 K to prevent condensation of solvent originating from droplets of the molten target. 

The MAPLE deposition procedure resulted in a deposition of a thin film of randomly-oriented, sub-1-$\mu$m-sized  diamonds on the glass coverslip, which was verified through optical and scanning electron microscopy (SEM). A more detailed characterization of the same diamond powder, including the size dispersion curve and SEM images of individual particles, can be found in our Ref.\cite{Filipkowski2022}. Importantly, the MAPLE-deposited film exhibits a fair homogeneity over the glass surface, however, with an insignificant fraction of particles found clustering. These agglomerates likely resulted from the clustering of nanocrystals in a powder and/or liquid suspension phase, rather than the deposition process itself. The adhesive properties of the deposited micro diamonds on the glass substrate are quite difficult to estimate. On the other hand, often silica substrates are used in NV diamond deposition techniques to enhance adhesion and avoid mobilization and crystal aggregation \cite{wood2022}. The well-established MAPLE deposition technique could practically be extended to any irregular surfaces without any experimental difficulty. Moreover, problem in wide-field imaging may arise due to randomly oriented diamonds being at different depth of focus of the objective as such not all fluorescent diamond spots can be well resolved. However, in practicality we observe that the ODMR signals are unaffected when the deposited diamond surface irregularity is within few micrometers.

\subsection*{The Experimental Setup}

Wide-field imaging magnetometry with NV centers in arbitrarily oriented diamond crystals was performed using the optical microscopy setup shown in Fig. \ref{fig:scheme}a. The wide-field microscopy setup consists primarily of a few optical elements, a microwave (MW) subsystem, and an imaging subsystem. The microwave subsystem (see Supplementary Fig. S3 online) comprises a microwave structure consisting of two straight copper striplines (100-$\mu$m wide) fabricated on a PCB board with a separation of 350 $\mu$m \cite{Mrozek2015} between them. One of the striplines transmits the microwave signal from a signal generator (Agilent N9310A). A Mini-Circuits ZRL-3500+ power amplifier is connected after the MW generator to amplify the microwave signal. The typical gain of the MW power amplifier is +17 dB at 2800 MHz. This microwave system is used to manipulate the ground-level spin states of the NV centers. The second stripline conducts a DC current that creates a local magnetic field to be characterized together with the bias field. MW and DC currents are applied simultaneously. The optical and imaging subsystem consists of a laser, dichroic mirror, lenses, microscope objective, and a camera for NV fluorescence imaging. The NV center ground-level spin triplet states were optically pumped to the excited state by a 530 nm green LED light source (Thorlabs M530L4). The pump beam was collimated by an aspheric condenser lens of focal length 50 mm and deflected by a dichroic mirror on the back focal plane of the 40X Olympus microscope objective with a numerical aperture of 0.65. The pump beam focused by the objective illuminates the top surface of the diamond sample over a field of view (FOV) of 308 x 246 $\mu$m. The red fluorescence emitted from the NV centers was collected using the same microscope objective used to deliver the pump beam. Additional optical filters and lenses ensure that the emitted red fluorescence was properly focused and imaged by an IDS UI-3240 CP camera equipped with a CMOS sensor with 12 bit depth. A small bias magnetic field (3.2 mT) from a neodymium permanent magnet is applied to lift the degeneracy between the states of m$_{s}$ = $\pm$ 1. 
\section*{Data availability}

The raw volumetric image datasets generated and analyzed during this study are available in the Zenodo data repository \cite{zenodo2022}.

\bibliography{ref.bib}

\section*{Acknowledgements}

The research was carried out within the TEAM NET programme of the Foundation for Polish Science co-financed by the European Union under the European Regional Development Fund, project POIR.04.04.00-00-1644/18. This research was funded in part by National Science Centre, Poland grant number 2020/39/I/ST3/02322. 
For the purpose of Open Access, the author has applied a CC-BY public copyright licence to any Author Accepted Manuscript (AAM) version arising from this submission.

\section*{Author contributions statement}

A.M.W conceived the experiment,  S.S. conducted the experiment, M.S. prepared the MAPLE samples, M.J.G and M.F. prepared diamond suspensions, S.S, M.M, W.G. and A.M.W analyzed the results. S.S wrote the manuscript.  All authors critically commented on the manuscript draft and approved the final version.

\section*{Competing interests}

The authors declare no competing interests.

\section*{Additional information}

\textbf{Correspondence} and requests for materials should be addressed to A.M.W. 


\end{document}


\flushbottom
\maketitle
\thispagestyle{empty}
\setlength{\abovedisplayskip}{3pt}
\setlength{\belowdisplayskip}{3pt}

\section*{\centering Supplementary information}  

\setcounter{figure}{0}
\renewcommand{\thefigure}{S\arabic{figure}}
\begin{figure}[H]
    \centering
    \includegraphics[scale=0.75]{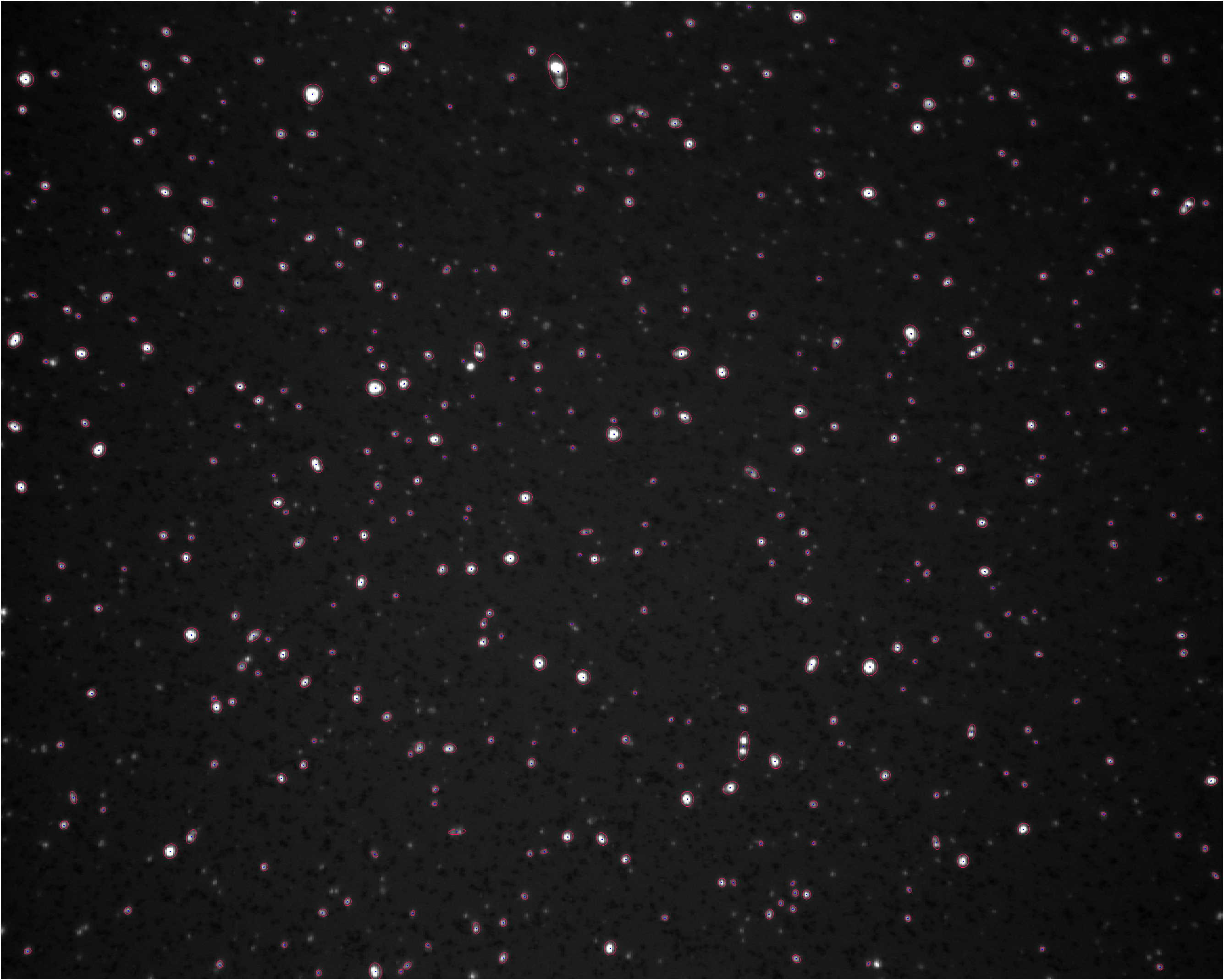}
    \caption{Bright fluorescent diamond spots identified by the automated ODMR reconstruction algorithm from a single image obtained by the camera. The pixel region is encircled in red with XY coordinates of the spots marked in blue.}
    \label{fig:f17}
\end{figure}
\begin{figure}[H]
    \centering
    \includegraphics[scale=1.2]{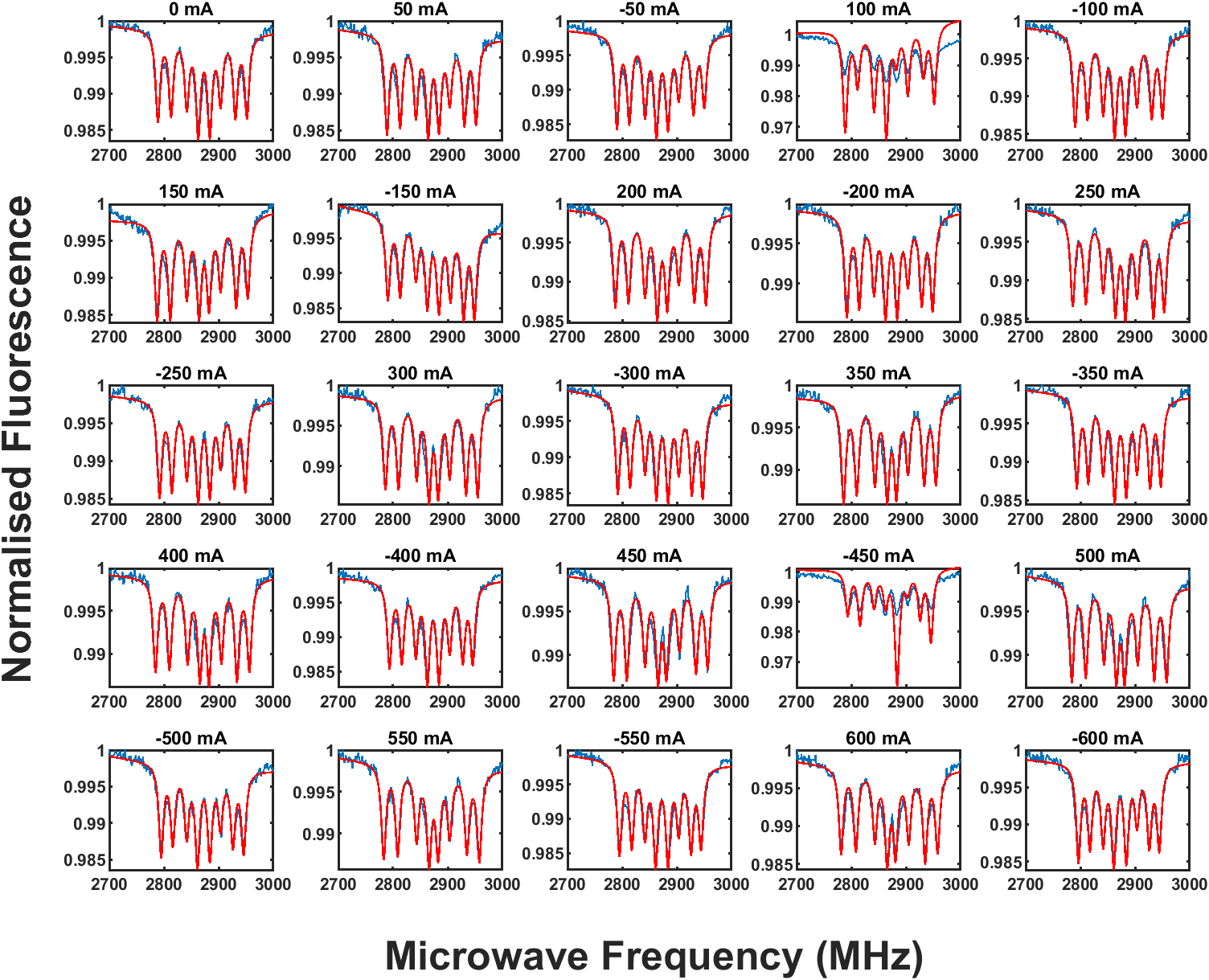}
    \caption{An example of multi-Lorentzian curve fitting to the ODMR spectra retrieved from a single diamond spot using the automated algorithm for different current values.}
    \label{fig:lorentzfit}
\end{figure}
\begin{figure}[H]
    \centering
    \includegraphics[scale=0.60]{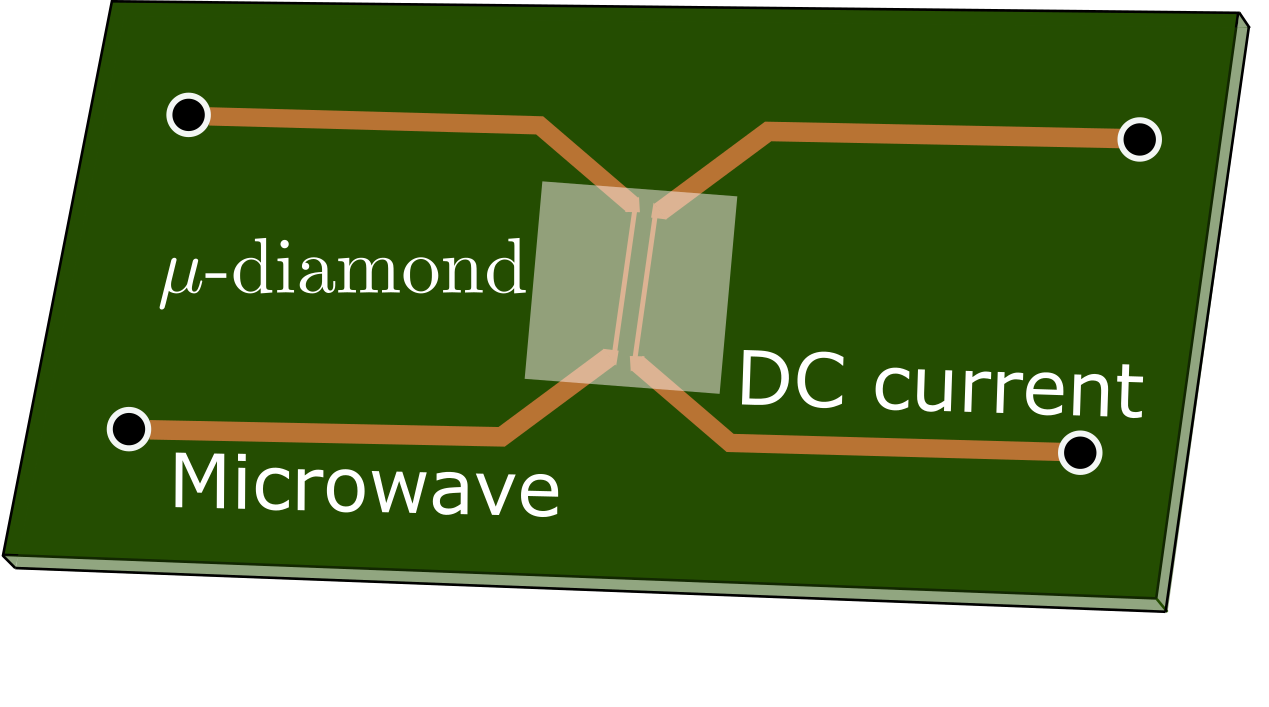}
    \caption{Schematic of the microwave and current striplines with a thin layer of micro-diamond powder deposited on a glass coverslip.}
    \label{fig:ant}
\end{figure}